\documentclass[times, twoside, watermark]{zHenriquesLab-StyleBioRxiv}

\leadauthor{Harju \& Broedersz}
\begin{document}

\title{Physical models of bacterial chromosomes}
\shorttitle{Minireview}

\author[1]{Janni Harju}
\author[1,2,\Letter]{Chase P. Brodersz}

\affil[1]{Department of Physics and Astronomy, Vrije Universiteit Amsterdam, 1081 HV Amsterdam, The Netherlands}
\affil[2]{Arnold Sommerfeld Center for Theoretical Physics and Center for NanoScience, Department of Physics, Ludwig-Maximilian-University Munich, Theresienstr. 37, D-80333 Munich, Germany}

\maketitle
\begin{abstract}
The interplay between bacterial chromosome organization and functions such as transcription and replication can be studied in increasing detail using novel experimental techniques. Interpreting the resulting quantitative data, however, can be theoretically challenging. In this minireview, we discuss how connecting experimental observations to biophysical theory and modeling can give rise to new insights on bacterial chromosome organization. We consider three flavors of models of increasing complexity: simple polymer models that explore how physical constraints, such as confinement or plectoneme branching, can affect bacterial chromosome organization; bottom-up mechanistic models that connect these constraints to their underlying causes, for instance chromosome compaction to macromolecular crowding, or supercoiling to transcription; and finally, data-driven methods for inferring interpretable and quantitative models directly from complex experimental data. Using recent examples, we discuss how biophysical models can both deepen our understanding of how bacterial chromosomes are structured, and give rise to novel predictions about bacterial chromosome organization.
\end {abstract}

\begin{corrauthor}
\href{mailto:c.p.brodersz@vu.nl}{c.p.broedersz \at vu.nl}
\end{corrauthor}

\section{Introduction}

The genome of many bacterial species is contained in a single circular chromosome, which is compressed by orders of magnitude into the bacterial cell. Microscopy studies have revealed that various factors, such as transcription, Nucleoid Associated Proteins (NAPs), supercoiling, loop-extrusion by SMC complexes, and replication shape bacterial chromosome organization~\citep{Dame2019,Lioy2021Oct, Yanez-Cuna2023Oct, Gogou2021}. However, since microscopy experiments cannot resolve the full 3D conformation of a bacterial chromosome, the effects of genetic and pharmocological perturbations on chromosome organization are often only explored indirectly, for instance by observing how they affect chromosome compaction or segregation.
Alternatively, chromosome organization can be studied using sequencing-based methods, including Chromosome Conformation Capture experiments such as Hi-C~\citep{Le2013}. Hi-C experiments measure how often pairs of loci are spatially proximate, or ``in contact'', averaged over a population of cells. Although interpreting Hi-C data remains challenging, these and other high-resolution quantitative data open new avenues to address old, yet unanswered questions: how are bacterial chromosomes organized across scales? how do various biological mechanisms control this organization? and, how does chromosome organization facilitate biological functions?

The recent surge in experimental techniques to quantitatively probe bacterial chromosome organization poses new and exciting challenges for biophysical modeling. Theoretical and computational biophysical models use concepts from polymer physics to explore how different mechanisms, such as macromolecular crowding~\citep{Rivas2016,Polson2018Aug} or bridging by NAPs~\citep{Dame2020Apr, Amemiya2021}, can affect chromosome organization and dynamics. This minireview focuses on recent models for bacterial chromosome organization, grouped according to their underlying modeling approach, and ordered by increasing complexity. First, we discuss polymer models that study \textit{how geometrical and topological constraints affect bacterial chromosomes}, such as how cellular confinement or polymer branching influence chromosome conformation and dynamics. Second, we explore bottom-up models, which study \textit{how chromosome organization emerges from microscopic mechanisms}, like how loop-extrusion by SMC complexes ($\sim 50$ nm)~\citep{Fudenberg2017Jan,Banigan2020} can organize chromosomes at the nucleoid scale ($\sim 1$ \textmu m). Finally, we consider data-driven approaches, which seek to  \textit{infer a model for chromosome organization, given experimental data}, providing a physical interpretation for Hi-C maps. We discuss various benefits and limitations of these different modeling approaches, explore future modeling opportunities and challenges, and summarize key insights into bacterial chromosome organization gained via biophysical modeling.

\section{Geometric and topological constraints determine chromosome organization}
\begin{figure*}
    \centering
    \includegraphics[width=\textwidth]{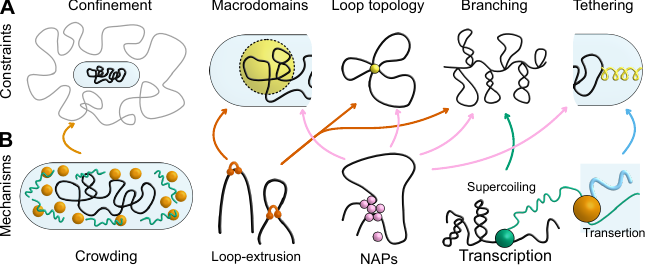}
    \caption{{\bf Constraint-based and bottom-up models.}
    \textbf{A} Examples of constraint-based models. Left to right: Polymer confinement can give rise to effects such as entropic segregation~\citep{Jun2006}. Constraining chromosomal regions corresponding to \textit{E. coli} macrodomains to subvolumes of the nucleoid can give rise to order and chromosome segregation~\citep{Junier2014Feb}. Fixed loop topologies can orient bacterial chromosomes and enhance their segregation~\citep{Mitra2022Nov}. Feather-boa models show that branching can affect chromosome compaction, segregation, and dynamics. Constraining origins of replication or other loci by tethering them to the cell membrane can give rise to linear ordering of the chromosome~\citep{Buenemann2010Nov, Buenemann2011May}.
    \textbf{B} Examples of bottom-up mechanisms. Left to right: Macromolecular crowding due to RNA, ribosomes, and other large molecules can compact bacterial chromosomes~\citep{Rivas2016}. Loop-extrusion by MukBEF can contribute to macrodomain formation in \textit{E. coli}~\citep{Lioy2018Feb}, whereas condensin can tie together the chromosomal arms in \textit{C. crescentus} and \textit{B. subtilis}~\citep{Le2013,Wang2017}. NAPs can impact chromosome organization by condensing macrodomains, by bridging, by stabilizing supercoils, or by associating with the cell membrane~\citep{Dame2020Apr, Amemiya2021}. Transcription gives rise to twin-supercoiled-domains, with positive supercoiling ahead and negative supercoiling behind RNA polymerases~\citep{Junier2023Oct}. The insertion of newly translated proteins into the cell membrane (transertion) causes certain genes to associate with the cell wall~\citep{Roggiani2015Nov}.
    }
    \label{fig:constraints}
\end{figure*}

In its essence, a bacterial chromosome is a long polymer confined to a small volume. The simplest physical models for bacterial chromosome organization explore how different, often biologically motivated, forces and constraints  affect the polymer's organization and dynamics (Fig.~\ref{fig:constraints}A).

\subsection{Geometric constraints}\label{sec:geometric_constraints}
The volume of confinement and constraining loci within the cell are examples of geometric constraints on a bacterial chromosome. Even these minimal constraints can explain well-known features of bacterial chromosome organization: tethering the origin of replication to a cell pole can give rise to linear chromosome organization~\citep{Buenemann2010Nov, Buenemann2011May}, as seen in species such as \textit{C. crescentus}, while tight confinement can drastically change the way that polymers interact, with implications for bacterial chromosome segregation.

\cite{Jun2006} proposed that bacterial chromosome segregation could be explained by entropic forces acting on two confined polymers. Consider two polymers consisting of $N$ monomers, confined to a long cylinder of diameter $d$. Since below length scales $d$, each polymer is unaffected by the confinement, we can split each polymer into subsections called ``confinement blobs'', inside which the polymer is unconstrained. Each blob is constrained to lie along the long axis of the cylinder, which introduces an entropic cost. The entropic cost of confinement can hence be shown to scale with the number of confinement blobs, $N_{\mathrm{blobs}}\sim Nd^{-1/\nu}$, where $\nu$ is the Flory exponent. Since overlap of any two blobs is entropically costly, the two polymers will entropically segregate.
Such arguments can be extended for ring polymers and for shorter cylinders~\citep{Jun2010Aug, Jung2012Jan}, suggesting that replicated bacterial chromosomes could entropically segregate in cellular confinement.

Despite theoretical arguments for entropic segregation, several simulation studies have shown that, without additional constraints, circular chromosomes do no segregate at intermediate replication stages. Constraints such as the concentric-shell model~\citep{Jun2006}, confining sections of the chromosome to sub-volumes of the nucleoid (modeling Macrodomains of the \textit{E. coli} chromosome)~\citep{Junier2014Feb}, fixing the replisomes at mid-cell~\citep{ElNajjar2020Jun}, or linking chromosomal arms with loop-extruders~\citep{Harju2023} have been necessary to achieve segregation concurrent with replication. Why are such additional constraints needed? We recently argued that at intermediate replication stages, purely entropic forces can actually \textit{inhibit} bacterial chromosome segregation by pushing replication forks apart~\citep{Harju2023}. Additionally, free energy calculations have shown that the time delay before entropic segregation begins can grow exponentially with the chain length~\citep{Minina2014, Minina2015Jul, Polson2018Aug}, and that two polymers of different lengths do not necessarily demix in confinement~\citep{Polson2021Jan}. Both due to partially conflicting simulation results and the lack of experimental evidence, the role of entropy in bacterial chromosome organization remains a subject of debate. Furthermore, we still lack theoretical understanding for chromosome segregation in spherically shaped cocci~\citep{Pinho2013}, and in species with multiple chromosomes of different topologies~\citep{Ren2022Feb}.

\subsection{Topological constraints}
The shape of a bacterial chromosome is characterized by topological constants: a linear, a circular, and a partially replicated chromosome all have different numbers of loops (0, 1, or 2) and are hence topologically distinct. We will now discuss how topological changes can affect the organization and dynamics of bacterial chromosomes.

\cite{Mitra2022July,Mitra2022Nov} recently proposed that fixed loops at the boundaries of \textit{E. coli} Macrodomains (Fig.~\ref{fig:constraints}A) could be sufficient to explain experimentally observed chromosome organization and segregation patterns. The authors showed that fixed loop architectures give rise to predictable and robust orientation of confined chromosomes, and that excluded volume interactions between loops can enhance chromosome segregation. Although it remains to be shown whether such stable loops at fixed genomic positions are common in bacteria, these findings also suggest that more randomly placed loops could affect the direction of entropic forces at the single-cell level.

Another example of a topological constraint on a bacterial chromosome is its supercoiling level. To illustrate, consider holding the ends of a piece of ribbon, and twisting them in opposite directions. This causes the ribbon to writhe around its central axis. If you now release tension by bringing the ends of the ribbon closer together, the ribbon will coil up into a plectoneme, but the number of turns (the linking number) will be conserved. Similarly, supercoiling of bacterial chromosomes by active mechanisms causes the DNA to branch into plectonemes~\citep{Dorman2019Dec, Junier2023Oct}. This observation motivated the development of ``feather-boa'' models, where the chromosome is considered to consist of loops or branches emanating from a backbone (Reviewed in~\cite{Ha2015}).

Branching can both compact chromosomes and enhance their segregation~\citep{Jun2010Aug}. Additionally, feather-boa models have been shown to reproduce experimentally observed features of bacterial chromosome organization and dynamics, such as subdiffusive motion of chromosomal loci~\citep{Yu2021}, and helical ordering of chromosomal arms~\citep{Swain2019}. Feather-boa models hence remain an active area of research, and new computational advances are improving simulation resolutions and speeds~\citep{Goodsell2018May,Ghobadpour2021Jul}. Despite these computational advances, an open-standing theoretical question is how branches and loops should be (dynamically) distributed in bacterial chromosome models, given what we know about their underlying causes.

\section{Bottom-up modeling}
In this section, we focus on bottom-up models, which model how chromosome organization emerges from proposed biological mechanisms (Fig.~\ref{fig:constraints}B). Such models can describe how transcription gives rise to plectoneme branches, or how macromolecular crowding confines the nucleoid to only 40-90\% of the cell~\citep{Gray2019}. 

A strength of bottom-up models is that they provide mechanistic insight and make novel predictions. A limitation is that more complex aspects of chromosome organization may be affected by several distinct mechanisms acting in unison. To illustrate, recent work by \cite{Joyeux2021Jan,Joyeux2023Feb} has shown that crowding and supercoiling can compact the chromosome in non-additive ways at high supercoiling densities, and that macromolecular crowding can enhance chromosome compaction by crosslinkers. These works illustrate that different bacterial chromosome organization mechanisms do not act in isolation. Despite such challenges, bottom-up models can provide conceptual insight into how various molecular mechanism control bacterial chromosome organization.

\subsection{Loop extrusion}
Hi-C experiments have revealed that bacterial chromosomes are more ordered than homogeneous, randomly oriented polymers in confinement. For instance, bacterial condensin mediates long-range contacts between the two chromosomal arms in species such as \textit{C. crescentus}~\citep{Le2013}, and \textit{B. subtilis}~\citep{Wang2017}, resulting in a prominent off-diagonal trace on Hi-C maps. Another SMC, MukBEF, on the other hand, enhances long-range contacts across large parts of the \textit{E. coli} chromosome~\citep{Lioy2018Feb}. These findings have triggered the development of loop extrusion models for bacterial chromosomes. 

One of the early questions addressed by biophysical modeling of SMCs in bacteria was whether these protein complexes move diffusively or via active loop extrusion. Whereas it was suggested that targeted loading of diffusive slip-links could be sufficient to model eukaryotic SMC behavior~\citep{Brackley2017Sep} and that MukBEF clustering could arise due to Turing patterning~\citep{Murray2017Oct}, a simulation model for \textit{B. subtilis}~\citep{Miermans2018} indicated that thousands of diffusive slip-links were needed to explain off-diagonal traces on bacterial Hi-C maps, while only tens of active loop-extruders were sufficient, more consistent with experimental reports of $\sim 30$ condensin complexes per chromosome~\citep{Wilhelm2015May}. These and other simulations, as well as mounting experimental evidence, have led to loop extrusion becoming more broadly accepted~\citep{Fudenberg2017Jan, Banigan2020}.

MukBEF loop extrusion in \textit{E. coli} has been modeled in 1D~\citep{Makela2020}. The authors proposed that non-targeted loading of MukBEF gives rise to an array of loops that spans most of the chromosome. Non-targeted loading of MukBEF could hence result in a feather-boa structure on parts of the \textit{E. coli} chromosome, reminiscent of microscopy observations of MukBEF distributions in widened \textit{E. coli} cells~\citep{Japaridze2023Mar}. However, future work still needs to address how loop extrusion by MukBEF could affect the 3D organization of \textit{E. coli} chromosomes.

The effects of loop extrusion on \textit{B. subtilis} chromosome organization, by contrast, have been modeled by combining 1D loop-extruder dynamics with 3D polymer simulations~\citep{Brandao2019,Brandao2021}.
These models better recapitulate Hi-C data if loop-extruders slow down as they collide with RNA polymerases at highly transcribed regions. Patterns on Hi-C maps for strains with two loop-extruder loading sites, on the other hand, can be explained if loop-extruders can traverse each other upon collision, as seen \textit{in vitro}~\citep{Kim2020Mar}.

More recently, we modeled how loop-extruders loaded at the origins of replication affect the segregation and organization of replicating bacterial chromosomes~\citep{Harju2023}. This so-called topo-entropic segregation model explains how the geometry and effective topology of a replicating chromosome affect the direction of entropic forces. We found that at intermediate replication stages, purely entropic forces inhibit bacterial chromosome segregation. However, loop-extruders loaded at the origins of replication effectively linearize partially replicated chromosomes, and this change in effective topology redirects entropic forces to enable concurrent replication and segregation.

\subsection{NAPs and phase separation}
\textit{In vitro} studies have shown that NAPs can locally twist, bend or bridge DNA~\citep{Song2015, Dame2020Apr,Amemiya2021}. This shows that NAPs can \textit{locally} affect DNA structure, but what is their impact on bacterial chromosome organization at larger scales? We will first discuss long-range bridging, which introduces transient cross-links on bacterial chromosomes. We then turn to liquid-liquid phase separation, which may allow compartmentalization within bacterial cells~\citep{Cohan2020Aug,Azaldegui2021Apr}.

Physically, NAPs can be modelled as particles that can diffuse, interact with each other, and bind to DNA. For bridging to occur, the number of DNA strands that the NAP can simultaneously bind to (its valency) should be at least two. \cite{Brackley2013Sep} showed that, even in the absence of NAP-NAP interactions or cooperative binding, multivalent binding could be sufficient to give rise to NAP clustering. A bivalently binding NAP introduces a loop on the chromosome, which is entropically costly. Two bivalent NAPs can bind far apart, giving rise to two loops, or next to each other, effectively giving rise to just one loop. Hence, even in the absence of NAP-NAP interactions, bridging proteins can cluster for entropic reasons.

Non-cooperative NAP binding can also affect chromosome dynamics; \cite{Subramanian2023Apr} showed that transient bridging can give rise to sub-diffusive motion of loci at timescales below the bridge lifetime. Consistent with this model, an H-NS mutant of \textit{E. coli} showed weaker sub-diffusive behavior of loci than the wild type.

Although some NAPs might bind non-cooperatively, many are known to interact, which can be modelled by introducing NAP-NAP interactions in simulations. \cite{Joyeux2021Jan} studied how protein self-association impacts chromosome organization. Inspired by the \textit{E. coli} NAP H-NS, two modes of NAP binding were modeled: filament- or cluster-forming. In simulations, filament-forming proteins stiffened chromosomal regions where they bound, but did not to compact DNA. Conversely, clustering proteins condensed the chromosome, but did not stiffen DNA. This work illustrates that even simple coarse-grained models can capture a variety of NAP behaviours.

Certain NAP-NAP interactions can give rise to collective phenomena, such as biomolecular condensation. For instance, HU and Dps, two important NAPs in \textit{E. coli}, have been observed to lead to phase separation of DNA segments \textit{in vitro}~\citep{Gupta2023May}. Put simply, phase separation can occur when attractive interactions start to dominate over entropic effects; whereas entropy favors spreading NAPs across the accessible volume, attractive NAP-NAP interactions of suitable geometry and sufficient range can favor NAP condensation. Since phase separation can create long-range order, it can impact chromosome organization at large scales.

Some of the earliest evidence for biomolecular condensation in bacterial cells came from observations of ParB clusters forming at \textit{parS} sites on bacterial chromosomes and plasmids~\citep{Broedersz2014Jun,Jalal2020Jun}. Bottom-up models have been used to explore transport~\citep{Lim2014May,Surovtsev2016Nov,Hu2017Apr,Walter2017Jul, Kohler2023Oct} and force generation~\citep{Hanauer2021Sep} by the ParAB\textit{S} system, as well as ParB cluster formation~\citep{Broedersz2014Jun, Sanchez2015Aug, Walter2021Apr}. These works offer two key physical insights. First, as opposed to earlier works that assumed that ParB only spreads along the one-dimensional DNA strand~\citep{Murray2006Sep, Breier2007May}, the formation of ParB clusters is an inherently three-dimensional phenomenon; a well-known result from statistical physics states that phase separation cannot occur in one-dimensional systems with short range interactions. Hence a combination of 1D spreading, 3D bridging, and fluctuations of the chromosome are important for the formation of ParB clusters. Accordingly, \textit{in vitro} experiments have confirmed that bridging is essential for ParB spreading~\citep{Graham2014May}, and that ParB-dimers can recruit each other in-trans and form dynamic clusters via bridging~\citep{Tisma2022Jun,Tisma2023Nov}. Second, the maintenance of separate ParB clusters consumes energy; to minimize their surface area, phase-separated droplets are expected to merge either via Ostwald ripening (when constituents diffuse from smaller droplets to larger ones) or by collision. This implies that the maintenance of ParB condensates on separate plasmids and/or \textit{parS} sites may require an active mechanism, such as ParA ATPase~\citep{Guilhas2020Jul} and/or ParB CTPase activity~\citep{Osorio-Valeriano2021Oct}.

\subsection{Effects of transcription}\label{sec:transcription}
Transcription and translation can affect bacterial chromosome organization in multiple ways. Steric interactions with ribosomes and RNA can affect nucleoid compaction and localization~\citep{Xiang2021Jul,Miangolarra2021Oct}. Transertion -- the insertion of membrane proteins into the cell wall as they are translated and transcribed~\citep{Roggiani2015Nov, Spahn2023Oct} -- can cause loci to remain near the cell membrane. Highly transcribed genes have been proposed to colocalize, since RNA polymerases can cluster in fast growth conditions~\citep{Ladouceur2020Aug}. Finally, transcription introduces both positive and negative supercoils, and highly transcribed genes can act as topological barriers that inhibit plectoneme diffusion~\citep{Le2013, Le2016}. Since single-molecule experiments are providing  evidence that some NAPs~\citep{Guo2021} and potentially SMCs~\citep{Kim2022Jul} are recruited to areas of high supercoiling, future models could explore the interplay of these different mechanisms of bacterial chromosome organization.

By comparing Monte Carlo simulations of a confined polymer to experimental data, \cite{Xiang2021Jul} showed that the mesh size of the \textit{E. coli} nucleoid is compatible with the chromosome being embedded in an effective poor solvent. Since ribosome and DNA densities were found to be anti-correlated, the authors suggested that this effective poor solvent could be a result of excluded volume interactions between the chromosome and ribosomes and/or RNA. \cite{Miangolarra2021Oct} further explored steric interactions between bacterial chromosomes and the transcriptional-translational machinery. By modeling the coupled 1D dynamics of DNA, ribosomes, and mRNA, they showed how active transcription and translation can affect the shape, size, and position of the nucleoid.

As reviewed by \cite{Junier2023Oct}, supercoiling due to transcription has not yet been modeled at scales of the bacterial chromosome. This is mainly due to computational limitations: bottom-up simulations for transcription-induced supercoiling in 3D have only been conducted for scales of tens of kilobases~\citep{Lepage2019Aug}. In light of these limitations, recent chromosome-scale models have considered branched polymers with plectoneme distributions that correlate with transcriptional activity~\citep{Hacker2017Jul,Wasim2023Apr, Wasim2023Jul}. To illustrate, \cite{Hacker2017Jul} divided the \textit{E. coli} chromosome into ``plectoneme-rich'' and ``plectoneme-free'' regions based on RNAP Chip-seq data, and then simulated branched polymers with sampled plectoneme configurations. Such use of complex, quantitative experimental data to constrain a model is a defining characteristic of modern data-driven modeling, which can offer new conceptual and mechanistic insights into bacterial chromosome organization.

\section{Data-driven models}
Over the last decade, Hi-C experiments have led to a breakthrough in studying chromosome organization quantitatively. A typical Hi-C map for a bacterial chromosome at a 5-10 kb resolution consists of $\sim 160 000$ data points, accurately probing features of chromosome organization over 3 orders of magnitude in genomic scales. Unlike microscopy methods, however, Hi-C experiments do not yield easily interpretable images, but rather a statistical metric for population-averaged pairwise contact counts. Using Hi-C data to faithfully extract information about the underlying distribution of three-dimensional chromosome configurations is thus a daunting theoretical challenge. 

Data-driven theoretical approaches seek to exploit the quantitative potential of Hi-C maps by directly inferring a model for 3D chromosome organization from experimental data~\citep{Contessoto2022}. Since Hi-C data represent an ensemble average of contact frequencies over the full distribution $P(\{{\bf r}\})$ of 3D chromosome conformations $\{{\bf r}\}$, data-driven models for bacterial chromosome organization usually seek to find either a single ``average'' chromosome consensus structure, $\{{\bf r}\}_{\mathrm{consensus}}$, or an ensemble of chromosome configurations, $P_{\mathrm{model}}(\{{\bf r}\})$~\citep{McCord2020}. Inference of both types of models is a technically challenging inverse problem, as we discuss below.

\begin{figure*}
    \includegraphics[width=\textwidth]{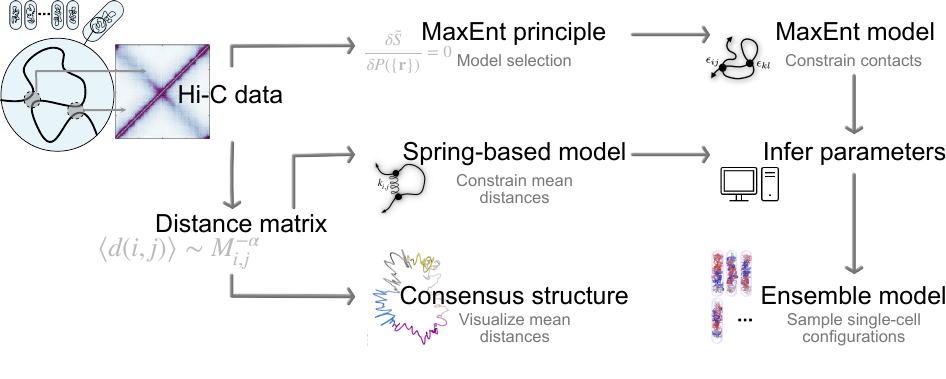}
    \caption{{\bf Data-driven modeling of bacterial chromosomes.} Data-driven models aim to infer a model for bacterial chromosomes from Hi-C data, which reflects the population-averaged contact counts between chromosomal regions. Subfigure shows Hi-C data for \textit{C. crescentus} swarmer cells~\citep{Le2013}. 
    Most models start by converting the Hi-C map to a distance matrix, for instance by assuming a scaling between mean distances between loci $d(i,j)$ and their contact counts $M_{i,j}$.
    Mean distances can be used to create a consensus structure, which depicts the estimated mean distances using a 3D curve. Subfigure shows consensus structure for \textit{E. coli}, adapted from~\citep{Hua2019}.
    Alternatively, mean distances can be used to constrain spring-based ensemble models, with effective harmonic potentials between loci.
    An ensemble model can also be inferred directly from Hi-C data: by maximizing the distribution entropy with constraints on contact probabilities ($\tilde S$), one can choose the least-assuming chromosome configuration distribution $P_{\mathrm{model}}(\{{\bf r}\})$ consistent with Hi-C data~\citep{Messelink2021}. The MaxEnt procedure selects a model that features effective close-range interactions between monomers.
    For both spring-based and MaxEnt ensemble models, the effective interaction parameters need to be inferred using computational approaches. 
    Once these parameters have been determined, the distribution can be sampled for single-cell chromosome configurations. Subfigure shows sampled configurations from the MaxEnt model for \textit{C. crescentus} swarmer cells~\citep{Messelink2021}.
    }
    \label{fig:data-driven}
\end{figure*}

\subsection{Consensus structure models}\label{sec:consensus_structure}
Most data-driven models don't use Hi-C data directly as input, and consensus structure algorithms are no exception (Fig.~\ref{fig:data-driven}). To construct a consensus structure, Hi-C scores are first converted into average spatial distances between locus pairs. This can be done by assuming that the mean distance between loci has a power-law scaling with the contact frequency~\citep{Marbouty2015}, or by using an experimentally determined calibration curve~\citep{Umbarger2011Oct}. Theoretically, however, the pairwise contact frequency between two foci is expected to depend not only on their mean distance, but also on the distance distribution's shape. Accordingly, experimental~\citep{Lioy2018Feb} and simulation~\citep{Messelink2021} results show that mean distances between chromosomal loci can show large deviations from average scalings.
Nevertheless, once an average distance map has been found, computational algorithms (reviewed by \cite{Liu2023Oct}) can be used to find a single 3D structure where the pairwise distances between loci are as compatible with the estimated mean distances as possible.

How should we interpret a consensus structure? Unlike proteins that often fold into specific, robust shapes that are critical for their function, bacterial chromosomes are highly flexible and dynamic polymers; imaging experiments show that the positions of chromosomal loci can vary by as much as half a cell length~\citep{Viollier2004Jun}. This inherent conformational variability is neglected by consensus structure algorithms: they cannot predict population-level variations in chromosome organization. Nonetheless, consensus structures may offer intuition for global chromosome organization by providing a ``convenient visualization tool''~\citep{Lioy2018Feb} for estimated mean distances between chromosomal regions. Furthermore, comparison of consensus structures for mutant strains or for drug-treated cells might yield clues about how different perturbations affect global chromosome organization.

\cite{Umbarger2011Oct} applied an algorithm originally developed for macromolecular assemblies such as nuclear pore complexes~\citep{Russel2012Jan} to predict consensus structures for a \textit{C. crescentus} chromosome based on 5C data. A set of candidate structures were found by initializing the algorithm with different initial conditions, and the inferred structures were then grouped by similarity. The model suggested that the arms of the \textit{C. crescentus} chromosome are wound in a loose helical structure. The authors also inferred structures for a mutant where the \textit{parS} site was relocated, leading to a shift in the cross-diagonal line on the 5C map. The corresponding consensus structure showed that the end of the nucleoid shifted to the new location of the \textit{parS} site, consistent with this site being tethered to a cell pole.

More recently, an error vector resultant algorithm was developed for faster and more accurate inference of consensus structures for prokaryotic chromosomes ~\citep{Hua2019}. The algorithm was applied to Hi-C data from \textit{C. crescentus}, \textit{E. coli} and \textit{B. subtilis}. By comparing consensus structures for wild-type and a $\Delta$\textit{fis} mutant of \textit{E. coli}, the authors concluded that the terminal region bends towards the rest of the chromosome in the mutant strain, reflecting increased Hi-C counts between the terminal region and the rest of the chromosome. Contrasting earlier consensus structures~\citep{Umbarger2011Oct, Marbouty2015}, helicity of the arms was only predicted for \textit{B. subtilis}. Such contradictory results raise further questions about how consensus structures relate to the underlying distribution of chromosome configurations in individual cells.

\subsection{Ensemble models}~\label{sec:ensemble}
Ensemble methods aim to capture population-level variability in bacterial chromosome organization by finding a \textit{distribution} $P_{\mathrm{model}}(\{{\bf r}\})$ of single-cell chromosome configurations, given Hi-C data (Fig.~\ref{fig:data-driven}). 
Most approaches assume an underlying statistical model for this distribution, defined by a set of effective interaction parameters. Once the effective parameters have been inferred from data, the model ensemble can be sampled using statistical methods such as Monte Carlo simulations. Samples from the distribution can be interpreted as single-cell chromosome configurations. In this way, an ensemble model constructed using population-averaged data can be used to make predictions about chromosome organization both on the single-cell and population level. 

Like consensus structure models, most data-driven ensemble models for bacterial chromosome organization start by assuming a relation between Hi-C scores and average monomer distances. However, these distances are now typically used to define spring-like interactions between loci, which constrain the mean distances in the model to match input data. For example, \cite{Yildirim2018May} constructed an ensemble model for the \textit{C. crescentus} chromosome by first converting Hi-C scores to expected distances between loci based on previous calibration data~\citep{Umbarger2011Oct}, and then constraining distances between a subset of monomer pairs in a plectonemic model using spring-like interactions. Equilibrium molecular dynamics simulations were used to produce an ensemble of chromosome configurations, and configurations were assigned statistical weights  based on how closely their distance matrices matched input Hi-C data. The correlation between the model's contact map and the experimental Hi-C map (0.88) was comparable to that between Hi-C maps of \textit{C. crescentus} and \textit{B. subtilis} (0.878 for Hi-C maps from~\citep{Le2013,Wang2015}). This illustrates that models constrained with inferred distances do not necessarily reproduce the Hi-C map faithfully.

Using a similar approach, \cite{Wasim2021} constructed a spring-based model for the \textit{E. coli} chromosome. They have since studied the sub-diffusional behavior of loci in their model, compared models for wild-type cells and HU- and MatP-mutants, and constructed a model with plectonemes~\citep{Bera2022, Wasim2023,Wasim2023Apr}. These works have suggested that locus (sub-)diffusion depends on genomic position, and that inclusion of plectonemes in the model slightly affected chromosome compaction, but not organization.

These and other ensemble techniques have advanced data-driven modeling beyond consensus structure inference for both pro- and eukaryotic chromosomes~\citep{Marti-Renom2018Oct,Oluwadare2019Dec,McCord2020}. However, several issues remain. Many ensemble approaches rely on strong assumptions, like thermal equilibrium or converting Hi-C counts to expected mean distances. Furthermore, the diversity of methods hints at a more fundamental concern: while all these approaches lead to \textit{a} model based on a given Hi-C map, many distinct ensembles could be consistent with the same data. So, how do you select the right one?

To address this challenge, our group developed a data-driven model for bacterial chromosome organization based on the Maximum Entropy (MaxEnt) principle. This principle selects the unique chromosome conformation distribution $P_{\mathrm{MaxEnt}}(\{{\bf r}\})$ that reproduces a given Hi-C map but is otherwise as unstructured as possible. Notably, the model does not rely on converting Hi-C scores to mean pairwise distances between loci. We applied this approach to model chromosome organization in new-born \textit{C. crescentus} swarmer cells~\citep{Messelink2021}. As validation, we showed that the model accurately predicts the long-axis distributions of loci over the entire chromosome, as measured by independent experiments~\citep{Viollier2004Jun}. The MaxEnt model can also reveal novel features of chromosome organization. For example, our model predicted the presence of ``super-domains'', or clusters of high chromosomal density at the single-cell level. The presence of these super-domains was validated using super-resolution microscopy. 

Our MaxEnt model was limited to new-born cells with a single chromosome, constrained using Hi-C data from synchronized cells~\citep{Le2013}. However, Hi-C experiments on \textit{E. coli}, \textit{B. subtilis}, and many other bacteria are conducted on asynchronous populations. The resulting Hi-C maps reflect an average over cells at different replication stages, which poses challenges for data-driven modeling. For instance, bacterial chromosome organization can vary over the cell cycle~\citep{Wang2014Sep}, and Hi-C experiments in replicating bacteria count both cis- and trans-contacts. \cite{Wasim2021} inferred models for \textit{E. coli} at discrete replication stages by constraining each model with the same asynchronous Hi-C data, and by assuming that trans-contacts are negligible. Since the validity of these approximations has yet to be established, an open-standing question in the field is how to best infer a model using Hi-C data from an asynchronous population. This is clearly a challenging but worthwhile problem: Such a model could provide new insight into the dynamics of chromosome organization across the cell cycle.

\section{Discussion and future challenges}
We have discussed three biophysical approaches to modeling bacterial chromosomes: models based on imposed geometric or topological constraints, bottom-up models, and data-driven approaches. Simple models of (branched) polymers in confinement have revealed how different constraints affect bacterial chromosome organization. Bottom-up approaches link these constraints to their underlying biophysical causes, and can hence be used to gain mechanistic insights. Data-driven methods aim to capture detailed chromosome organization by inferring models from experimental Hi-C data. Importantly, we note that these modeling approaches can complement each other. For instance, data-driven approaches can help hypothesize simple physical principles of chromosome organization, which can then guide the construction of bottom-up or constraint-based models.

Given that simplified constraint-based and bottom-up models are tailored to explain only certain aspects of chromosome organization, it can be difficult to test their predictions \textit{in vivo} with controlled and targeted perturbations. To test whether effects seen in polymer simulations are relevant at biological length- and time-scales, these simplified models could be compared to artificial systems of chromosomes in confinement~\citep{Birnie2021}. Conversely, bottom-up models could be constructed for ``simpler'' organisms: \cite{Stevens2023} recently presented a model for an entire, minimal bacterial cell, with a 543 kb long circular chromosome. For more complex systems, the increased availability of high-quality quantitative data creates opportunities for data-driven modeling. For instance, multiple types of data, such as Hi-C, imaging, and/or RNA-sequencing, can be combined to infer data-driven models that capture bacterial chromosome organization in its full complexity~\citep{Messelink2021,Wasim2023Apr}.

In conclusion, biophysical modeling has helped shape our understanding of how bacterial chromosomes are functionally organized. Since biophysical models can be easily adapted for different organisms, modeling can help search for divergent and unifying principles of prokaryotic genome organization.

\begin{acknowledgements}
We thank the members of our group for helpful discussions, and the organizers and attendants of the 2023 "Biology and physics of the prokaryotic chromosome" workshop at the Lorentz Center in Leiden, the Netherlands.
\end{acknowledgements}

\section*{Bibliography}
\bibliography{sources}

\begin{thebibliography}{101}
\providecommand{\natexlab}[1]{#1}
\providecommand{\url}[1]{\texttt{#1}}
\expandafter\ifx\csname urlstyle\endcsname\relax
  \providecommand{\doi}[1]{doi: #1}\else
  \providecommand{\doi}{doi: \begingroup \urlstyle{rm}\Url}\fi

\bibitem[Amemiya et~al.(2021)Amemiya, Schroeder, and Freddolino]{Amemiya2021}
Haley~M. Amemiya, Jeremy Schroeder, and Peter~L. Freddolino.
\newblock {Nucleoid-associated proteins shape chromatin structure and transcriptional regulation across the bacterial kingdom}.
\newblock \emph{Transcription}, 12\penalty0 (4):\penalty0 182, 2021.
\newblock \doi{10.1080/21541264.2021.1973865}.

\bibitem[Azaldegui et~al.(2021)Azaldegui, Vecchiarelli, and Biteen]{Azaldegui2021Apr}
Christopher~A. Azaldegui, Anthony~G. Vecchiarelli, and Julie~S. Biteen.
\newblock {The emergence of phase separation as an organizing principle in bacteria}.
\newblock \emph{Biophys. J.}, 120\penalty0 (7):\penalty0 1123--1138, April 2021.
\newblock ISSN 0006-3495.
\newblock \doi{10.1016/j.bpj.2020.09.023}.

\bibitem[Banigan and Mirny(2020)]{Banigan2020}
Edward~J. Banigan and Leonid~A. Mirny.
\newblock Loop extrusion: theory meets single-molecule experiments.
\newblock \emph{Current Opinion in Cell Biology}, 64:\penalty0 124--138, 6 2020.
\newblock ISSN 0955-0674.
\newblock \doi{10.1016/J.CEB.2020.04.011}.

\bibitem[Bera et~al.(2022)Bera, Wasim, and Mondal]{Bera2022}
Palash Bera, Abdul Wasim, and Jagannath Mondal.
\newblock { Hi-C embedded polymer model of Escherichia coli reveals the origin of heterogeneous subdiffusion in chromosomal loci }.
\newblock \emph{Physical Review E}, 105:\penalty0 064402, 6 2022.
\newblock ISSN 2470-0045.
\newblock \doi{10.1103/PhysRevE.105.064402}.

\bibitem[Birnie and Dekker(2021)]{Birnie2021}
Anthony Birnie and Cees Dekker.
\newblock {Genome-in-a-Box:}.
\newblock \emph{ACS Nano}, 15\penalty0 (1):\penalty0 111, January 2021.
\newblock \doi{10.1021/acsnano.0c07397}.

\bibitem[Brackley et~al.(2017)Brackley, Johnson, Michieletto, Morozov, Nicodemi, Cook, and Marenduzzo]{Brackley2017Sep}
C.~A. Brackley, J.~Johnson, D.~Michieletto, A.~N. Morozov, M.~Nicodemi, P.~R. Cook, and D.~Marenduzzo.
\newblock {Nonequilibrium Chromosome Looping via Molecular Slip Links}.
\newblock \emph{Phys. Rev. Lett.}, 119\penalty0 (13):\penalty0 138101, September 2017.
\newblock \doi{10.1103/PhysRevLett.119.138101}.

\bibitem[Brackley et~al.(2013)Brackley, Taylor, Papantonis, Cook, and Marenduzzo]{Brackley2013Sep}
Chris~A. Brackley, Stephen Taylor, Argyris Papantonis, Peter~R. Cook, and Davide Marenduzzo.
\newblock {Nonspecific bridging-induced attraction drives clustering of DNA-binding proteins and genome organization}.
\newblock \emph{Proc. Natl. Acad. Sci. U.S.A.}, 110\penalty0 (38):\penalty0 E3605--E3611, September 2013.
\newblock \doi{10.1073/pnas.1302950110}.

\bibitem[Brand{\ifmmode\tilde{a}\else\~{a}\fi}o et~al.(2019)Brand{\ifmmode\tilde{a}\else\~{a}\fi}o, Paul, van~den Berg, Rudner, Wang, and Mirny]{Brandao2019}
Hugo~B. Brand{\ifmmode\tilde{a}\else\~{a}\fi}o, Payel Paul, Aafke~A. van~den Berg, David~Z. Rudner, Xindan Wang, and Leonid~A. Mirny.
\newblock {RNA polymerases as moving barriers to condensin loop extrusion}.
\newblock \emph{Proc. Natl. Acad. Sci. U.S.A.}, 116\penalty0 (41):\penalty0 20489--20499, October 2019.
\newblock \doi{10.1073/pnas.1907009116}.

\bibitem[Brand{\ifmmode\tilde{a}\else\~{a}\fi}o et~al.(2021)Brand{\ifmmode\tilde{a}\else\~{a}\fi}o, Ren, Karaboja, Mirny, and Wang]{Brandao2021}
Hugo~B. Brand{\ifmmode\tilde{a}\else\~{a}\fi}o, Zhongqing Ren, Xheni Karaboja, Leonid~A. Mirny, and Xindan Wang.
\newblock {DNA-loop-extruding SMC complexes can traverse one another in vivo}.
\newblock \emph{Nat. Struct. Mol. Biol.}, 28:\penalty0 642--651, August 2021.
\newblock ISSN 1545-9985.
\newblock \doi{10.1038/s41594-021-00626-1}.

\bibitem[Breier and Grossman(2007)]{Breier2007May}
Adam~M. Breier and Alan~D. Grossman.
\newblock {Whole-genome analysis of the chromosome partitioning and sporulation protein Spo0J (ParB) reveals spreading and origin-distal sites on the Bacillus subtilis chromosome}.
\newblock \emph{Mol. Microbiol.}, 64\penalty0 (3):\penalty0 703--718, May 2007.
\newblock ISSN 0950-382X.
\newblock \doi{10.1111/j.1365-2958.2007.05690.x}.

\bibitem[Broedersz et~al.(2014)Broedersz, Wang, Meir, Loparo, Rudner, and Wingreen]{Broedersz2014Jun}
Chase~P. Broedersz, Xindan Wang, Yigal Meir, Joseph~J. Loparo, David~Z. Rudner, and Ned~S. Wingreen.
\newblock {Condensation and localization of the partitioning protein ParB on the bacterial chromosome}.
\newblock \emph{Proc. Natl. Acad. Sci. U.S.A.}, 111\penalty0 (24):\penalty0 8809--8814, June 2014.
\newblock \doi{10.1073/pnas.1402529111}.

\bibitem[Buenemann and Lenz(2010)]{Buenemann2010Nov}
Mathias Buenemann and Peter Lenz.
\newblock {A Geometrical Model for DNA Organization in Bacteria}.
\newblock \emph{PLoS One}, 5\penalty0 (11):\penalty0 e13806, November 2010.
\newblock ISSN 1932-6203.
\newblock \doi{10.1371/journal.pone.0013806}.

\bibitem[Buenemann and Lenz(2011)]{Buenemann2011May}
Mathias Buenemann and Peter Lenz.
\newblock {Geometrical ordering of DNA in bacteria}.
\newblock \emph{Commun. Integr. Biol.}, pages 291--293, May 2011.
\newblock ISSN 1091-4891.

\bibitem[Cohan and Pappu(2020)]{Cohan2020Aug}
Megan~C. Cohan and Rohit~V. Pappu.
\newblock {Making the Case for Disordered Proteins and Biomolecular Condensates in Bacteria}.
\newblock \emph{Trends Biochem. Sci.}, 45\penalty0 (8):\penalty0 668--680, August 2020.
\newblock ISSN 0968-0004.
\newblock \doi{10.1016/j.tibs.2020.04.011}.

\bibitem[Contessoto et~al.(2022)Contessoto, Cheng, and Onuchic]{Contessoto2022}
Vinícius~G. Contessoto, Ryan~R. Cheng, and José~N. Onuchic.
\newblock { Uncovering the statistical physics of 3D chromosomal organization using data-driven modeling }.
\newblock \emph{Current Opinion in Structural Biology}, 75:\penalty0 102418, 8 2022.
\newblock ISSN 0959-440X.
\newblock \doi{10.1016/J.SBI.2022.102418}.

\bibitem[Dame et~al.(2019)Dame, Rashid, and Grainger]{Dame2019}
Remus~T. Dame, Fatema Zahra~M. Rashid, and David~C. Grainger.
\newblock Chromosome organization in bacteria: mechanistic insights into genome structure and function.
\newblock \emph{Nature Reviews Genetics 2019 21:4}, 21:\penalty0 227--242, 11 2019.
\newblock ISSN 1471-0064.
\newblock \doi{10.1038/s41576-019-0185-4}.

\bibitem[Dame et~al.(2020)Dame, Rashid, and Grainger]{Dame2020Apr}
Remus~T. Dame, Fatema-Zahra~M. Rashid, and David~C. Grainger.
\newblock {Chromosome organization in bacteria: mechanistic insights into genome structure and function}.
\newblock \emph{Nat. Rev. Genet.}, 21:\penalty0 227--242, April 2020.
\newblock ISSN 1471-0064.
\newblock \doi{10.1038/s41576-019-0185-4}.

\bibitem[Dorman(2019)]{Dorman2019Dec}
Charles~J. Dorman.
\newblock {DNA supercoiling and transcription in bacteria: a two-way street}.
\newblock \emph{BMC Mol. Cell Biol.}, 20\penalty0 (1):\penalty0 1--9, December 2019.
\newblock ISSN 2661-8850.
\newblock \doi{10.1186/s12860-019-0211-6}.

\bibitem[El~Najjar et~al.(2020)El~Najjar, Geisel, Schmidt, Dersch, Mayer, Hartmann, Eckhardt, Lenz, and P.~L.]{ElNajjar2020Jun}
N.~El~Najjar, D.~Geisel, F.~Schmidt, S.~Dersch, B.~Mayer, R.~Hartmann, B.~Eckhardt, P.~Lenz, and Graumann P.~L.
\newblock {Chromosome Segregation in Bacillus subtilis Follows an Overall Pattern of Linear Movement and Is Highly Robust against Cell Cycle Perturbations.}
\newblock \emph{mSphere}, 5\penalty0 (3):\penalty0 e00255--20, June 2020.
\newblock ISSN 2379-5042.
\newblock \doi{10.1128/msphere.00255-20}.

\bibitem[Fudenberg et~al.(2017)Fudenberg, Abdennur, Imakaev, Goloborodko, and L.~A.]{Fudenberg2017Jan}
G.~Fudenberg, N.~Abdennur, M.~Imakaev, A.~Goloborodko, and Mirny L.~A.
\newblock {Emerging Evidence of Chromosome Folding by Loop Extrusion.}
\newblock \emph{Cold Spring Harbor Symp. Quant. Biol.}, 82:\penalty0 45--55, January 2017.
\newblock ISSN 0091-7451.
\newblock \doi{10.1101/sqb.2017.82.034710}.

\bibitem[Ghobadpour et~al.(2021)Ghobadpour, Kolb, Ejtehadi, and Everaers]{Ghobadpour2021Jul}
Elham Ghobadpour, Max Kolb, Mohammad~Reza Ejtehadi, and Ralf Everaers.
\newblock {Monte Carlo simulation of a lattice model for the dynamics of randomly branching double-folded ring polymers}.
\newblock \emph{Phys. Rev. E}, 104\penalty0 (1):\penalty0 014501, July 2021.
\newblock ISSN 2470-0053.
\newblock \doi{10.1103/PhysRevE.104.014501}.

\bibitem[Gogou et~al.(2021)Gogou, Japaridze, and Dekker]{Gogou2021}
Christos Gogou, Aleksandre Japaridze, and Cees Dekker.
\newblock {Mechanisms for Chromosome Segregation in Bacteria}.
\newblock \emph{Front. Microbiol.}, 12:\penalty0 685687, June 2021.
\newblock ISSN 1664-302X.
\newblock \doi{10.3389/fmicb.2021.685687}.

\bibitem[Goodsell et~al.(2018)Goodsell, Autin, and Olson]{Goodsell2018May}
David~S. Goodsell, Ludovic Autin, and Arthur~J. Olson.
\newblock {Lattice Models of Bacterial Nucleoids}.
\newblock \emph{J. Phys. Chem. B}, 122\penalty0 (21):\penalty0 5441--5447, May 2018.
\newblock ISSN 1520-6106.
\newblock \doi{10.1021/acs.jpcb.7b11770}.

\bibitem[Graham et~al.(2014)Graham, Wang, Song, Etson, van Oijen, Rudner, and Loparo]{Graham2014May}
Thomas G.~W. Graham, Xindan Wang, Dan Song, Candice~M. Etson, Antoine~M. van Oijen, David~Z. Rudner, and Joseph~J. Loparo.
\newblock {ParB spreading requires DNA bridging}.
\newblock \emph{Genes Dev.}, 28\penalty0 (11):\penalty0 1228--1238, May 2014.
\newblock ISSN 0890-9369.
\newblock \doi{10.1101/gad.242206.114}.

\bibitem[Gray et~al.(2019)Gray, Govers, Xiang, Parry, Campos, Kim, and Jacobs-Wagner]{Gray2019}
William~T. Gray, Sander~K. Govers, Yingjie Xiang, Bradley~R. Parry, Manuel Campos, Sangjin Kim, and Christine Jacobs-Wagner.
\newblock Nucleoid size scaling and intracellular organization of translation across bacteria.
\newblock \emph{Cell}, 177:\penalty0 1632, 5 2019.
\newblock ISSN 10974172.
\newblock \doi{10.1016/J.CELL.2019.05.017}.

\bibitem[Guilhas et~al.(2020)Guilhas, Walter, Rech, David, Walliser, Palmeri, Mathieu-Demaziere, Parmeggiani, Bouet, Le~Gall, and Nollmann]{Guilhas2020Jul}
Baptiste Guilhas, Jean-Charles Walter, Jerome Rech, Gabriel David, Nils~Ole Walliser, John Palmeri, Celine Mathieu-Demaziere, Andrea Parmeggiani, Jean-Yves Bouet, Antoine Le~Gall, and Marcelo Nollmann.
\newblock {ATP-Driven Separation of Liquid Phase Condensates in Bacteria}.
\newblock \emph{Mol. Cell}, 79\penalty0 (2):\penalty0 293--303.e4, July 2020.
\newblock ISSN 1097-2765.
\newblock \doi{10.1016/j.molcel.2020.06.034}.

\bibitem[Guo et~al.(2021)Guo, Kawamura, Littlehale, Marko, and Laub]{Guo2021}
Monica~S. Guo, Ryo Kawamura, Megan Littlehale, John~F. Marko, and Michael~T. Laub.
\newblock High-resolution, genome-wide mapping of positive supercoiling in chromosomes.
\newblock \emph{eLife}, 10, 7 2021.
\newblock ISSN 2050084X.
\newblock \doi{10.7554/ELIFE.67236}.

\bibitem[Gupta et~al.(2023)Gupta, Joshi, Arora, Mukhopadhyay, and Guptasarma]{Gupta2023May}
Archit Gupta, Ashish Joshi, Kanika Arora, Samrat Mukhopadhyay, and Purnananda Guptasarma.
\newblock {The bacterial nucleoid-associated proteins, HU and Dps, condense DNA into context-dependent biphasic or multiphasic complex coacervates}.
\newblock \emph{J. Biol. Chem.}, 299\penalty0 (5), May 2023.
\newblock ISSN 0021-9258.
\newblock \doi{10.1016/j.jbc.2023.104637}.

\bibitem[Ha and Jung(2015)]{Ha2015}
Bae~Yeun Ha and Youngkyun Jung.
\newblock Polymers under confinement: single polymers, how they interact, and as model chromosomes.
\newblock \emph{Soft Matter}, 11:\penalty0 2333--2352, 3 2015.
\newblock ISSN 17446848.
\newblock \doi{10.1039/C4SM02734E}.

\bibitem[Hacker et~al.(2017)Hacker, Li, and Elcock]{Hacker2017Jul}
William~C. Hacker, Shuxiang Li, and Adrian~H. Elcock.
\newblock {Features of genomic organization in a nucleotide-resolution molecular model of the Escherichia coli chromosome}.
\newblock \emph{Nucleic Acids Res.}, 45\penalty0 (13):\penalty0 7541--7554, July 2017.
\newblock ISSN 0305-1048.
\newblock \doi{10.1093/nar/gkx541}.

\bibitem[Hanauer et~al.(2021)Hanauer, Bergeler, Frey, and Broedersz]{Hanauer2021Sep}
Christian Hanauer, Silke Bergeler, Erwin Frey, and Chase~P. Broedersz.
\newblock {Theory of Active Intracellular Transport by DNA Relaying}.
\newblock \emph{Phys. Rev. Lett.}, 127\penalty0 (13):\penalty0 138101, September 2021.
\newblock ISSN 1079-7114.
\newblock \doi{10.1103/PhysRevLett.127.138101}.

\bibitem[Harju et~al.(2023)Harju, van Teeseling, and Broedersz]{Harju2023}
Janni Harju, Muriel C.~F. van Teeseling, and Chase~P. Broedersz.
\newblock Loop-extruders alter bacterial chromosome topology to direct entropic forces for segregation.
\newblock \emph{bioRxiv}, page 2023.06.30.547230, 7 2023.
\newblock \doi{10.1101/2023.06.30.547230}.

\bibitem[Hu et~al.(2017)Hu, Vecchiarelli, Mizuuchi, Neuman, and Liu]{Hu2017Apr}
Longhua Hu, Anthony~G. Vecchiarelli, Kiyoshi Mizuuchi, Keir~C. Neuman, and Jian Liu.
\newblock {Brownian Ratchet Mechanism for Faithful Segregation of Low-Copy-Number Plasmids}.
\newblock \emph{Biophys. J.}, 112\penalty0 (7):\penalty0 1489--1502, April 2017.
\newblock ISSN 0006-3495.
\newblock \doi{10.1016/j.bpj.2017.02.039}.

\bibitem[Hua and Ma(2019)]{Hua2019}
Kang-Jian Hua and Bin-Guang Ma.
\newblock {EVR: reconstruction of bacterial chromosome 3D structure models using error-vector resultant algorithm}.
\newblock \emph{BMC Genomics}, 20\penalty0 (1):\penalty0 738., October 2019.
\newblock ISSN 1471-2164.
\newblock \doi{10.1186/s12864-019-6096-0}.

\bibitem[Jalal and Le(2020)]{Jalal2020Jun}
Adam S.~B. Jalal and Tung B.~K. Le.
\newblock {Bacterial chromosome segregation by the ParABS system}.
\newblock \emph{Open Biol.}, 10\penalty0 (6):\penalty0 200097, June 2020.
\newblock ISSN 2046-2441.
\newblock \doi{10.1098/rsob.200097}.

\bibitem[Japaridze et~al.(2023)Japaridze, van Wee, Gogou, Kerssemakers, van~den Berg, and Dekker]{Japaridze2023Mar}
Aleksandre Japaridze, Raman van Wee, Christos Gogou, Jacob W.~J. Kerssemakers, Daan~F. van~den Berg, and Cees Dekker.
\newblock {MukBEF-dependent chromosomal organization in widened Escherichia coli}.
\newblock \emph{Front. Microbiol.}, 14:\penalty0 1107093, March 2023.
\newblock ISSN 1664-302X.
\newblock \doi{10.3389/fmicb.2023.1107093}.

\bibitem[Joyeux(2021)]{Joyeux2021Jan}
Marc Joyeux.
\newblock {Impact of Self-Association on the Architectural Properties of Bacterial Nucleoid Proteins}.
\newblock \emph{Biophys. J.}, 120\penalty0 (2):\penalty0 370--378, January 2021.
\newblock ISSN 0006-3495.
\newblock \doi{10.1016/j.bpj.2020.12.006}.

\bibitem[Joyeux(2023)]{Joyeux2023Feb}
Marc Joyeux.
\newblock {Organization of the bacterial nucleoid by DNA-bridging proteins and globular crowders}.
\newblock \emph{Front. Microbiol.}, 14:\penalty0 1116776, February 2023.
\newblock ISSN 1664-302X.
\newblock \doi{10.3389/fmicb.2023.1116776}.

\bibitem[Jun and Mulder(2006)]{Jun2006}
Suckjoon Jun and Bela Mulder.
\newblock {Entropy-driven spatial organization of highly confined polymers: Lessons for the bacterial chromosome}.
\newblock \emph{Proc. Natl. Acad. Sci. U.S.A.}, 103\penalty0 (33):\penalty0 12388--12393, August 2006.
\newblock \doi{10.1073/pnas.0605305103}.

\bibitem[Jun and Wright(2010)]{Jun2010Aug}
Suckjoon Jun and Andrew Wright.
\newblock {Entropy as the driver of chromosome segregation}.
\newblock \emph{Nat. Rev. Microbiol.}, 8:\penalty0 600--607, August 2010.
\newblock ISSN 1740-1534.
\newblock \doi{10.1038/nrmicro2391}.

\bibitem[Jung et~al.(2012)Jung, Jeon, Kim, Jeong, Jun, and Ha]{Jung2012Jan}
Youngkyun Jung, Chanil Jeon, Juin Kim, Hawoong Jeong, Suckjoon Jun, and Bae-Yeun Ha.
\newblock {Ring polymers as model bacterial chromosomes: confinement, chain topology, single chain statistics, and how they interact}.
\newblock \emph{Soft Matter}, 8\penalty0 (7):\penalty0 2095--2102, January 2012.
\newblock ISSN 1744-683X.
\newblock \doi{10.1039/C1SM05706E}.

\bibitem[Junier et~al.(2014)Junier, Boccard, and Esp{\ifmmode\acute{e}\else\'{e}\fi}li]{Junier2014Feb}
Ivan Junier, Fr{\ifmmode\acute{e}\else\'{e}\fi}d{\ifmmode\acute{e}\else\'{e}\fi}ric Boccard, and Olivier Esp{\ifmmode\acute{e}\else\'{e}\fi}li.
\newblock {Polymer modeling of the E. coli genome reveals the involvement of locus positioning and macrodomain structuring for the control of chromosome conformation and segregation}.
\newblock \emph{Nucleic Acids Res.}, 42\penalty0 (3):\penalty0 1461--1473, February 2014.
\newblock ISSN 0305-1048.
\newblock \doi{10.1093/nar/gkt1005}.

\bibitem[Junier et~al.(2023)Junier, Ghobadpour, Espeli, and Everaers]{Junier2023Oct}
Ivan Junier, Elham Ghobadpour, Olivier Espeli, and Ralf Everaers.
\newblock {DNA supercoiling in bacteria: state of play and challenges from a viewpoint of physics based modeling}.
\newblock \emph{Front. Microbiol.}, 14:\penalty0 1192831, October 2023.
\newblock ISSN 1664-302X.
\newblock \doi{10.3389/fmicb.2023.1192831}.

\bibitem[Kim et~al.(2020)Kim, Kerssemakers, Shaltiel, Haering, and Dekker]{Kim2020Mar}
Eugene Kim, Jacob Kerssemakers, Indra~A. Shaltiel, Christian~H. Haering, and Cees Dekker.
\newblock {DNA-loop extruding condensin complexes can traverse one another}.
\newblock \emph{Nature}, 579:\penalty0 438--442, March 2020.
\newblock ISSN 1476-4687.
\newblock \doi{10.1038/s41586-020-2067-5}.

\bibitem[Kim et~al.(2022)Kim, Gonzalez, Pradhan, van~der Torre, and Dekker]{Kim2022Jul}
Eugene Kim, Alejandro~Martin Gonzalez, Biswajit Pradhan, Jaco van~der Torre, and Cees Dekker.
\newblock {Condensin-driven loop extrusion on supercoiled DNA}.
\newblock \emph{Nat. Struct. Mol. Biol.}, 29:\penalty0 719--727, July 2022.
\newblock ISSN 1545-9985.
\newblock \doi{10.1038/s41594-022-00802-x}.

\bibitem[K{\ifmmode\ddot{o}\else\"{o}\fi}hler and Murray(2023)]{Kohler2023Oct}
Robin K{\ifmmode\ddot{o}\else\"{o}\fi}hler and Se{\ifmmode\acute{a}\else\'{a}\fi}n~M. Murray.
\newblock {Putting the Par back into ParABS: Plasmid Partitioning Driven by ParA Oscillations}.
\newblock \emph{bioRxiv}, page 2023.10.16.562490, October 2023.
\newblock URL \url{https://doi.org/10.1101/2023.10.16.562490}.

\bibitem[Ladouceur et~al.(2020)Ladouceur, Parmar, Biedzinski, Wall, Tope, Cohn, Kim, Soubry, Reyes-Lamothe, and Weber]{Ladouceur2020Aug}
Anne-Marie Ladouceur, Baljyot~Singh Parmar, Stefan Biedzinski, James Wall, S.~Graydon Tope, David Cohn, Albright Kim, Nicolas Soubry, Rodrigo Reyes-Lamothe, and Stephanie~C. Weber.
\newblock {Clusters of bacterial RNA polymerase are biomolecular condensates that assemble through liquid{\textendash}liquid phase separation}.
\newblock \emph{Proc. Natl. Acad. Sci. U.S.A.}, 117\penalty0 (31):\penalty0 18540--18549, August 2020.
\newblock \doi{10.1073/pnas.2005019117}.

\bibitem[Le et~al.(2013)Le, Imakaev, Mirny, and Laub]{Le2013}
Tung B.~K. Le, Maxim~V. Imakaev, Leonid~A. Mirny, and Michael~T. Laub.
\newblock {High-Resolution Mapping of the Spatial Organization of a Bacterial Chromosome}.
\newblock \emph{Science}, 342\penalty0 (6159):\penalty0 731--734, November 2013.
\newblock ISSN 0036-8075.
\newblock \doi{10.1126/science.1242059}.

\bibitem[Le and Laub(2016)]{Le2016}
Tung~BK Le and Michael~T Laub.
\newblock Transcription rate and transcript length drive formation of chromosomal interaction domain boundaries.
\newblock \emph{The EMBO Journal}, 35:\penalty0 1582--1595, 7 2016.
\newblock ISSN 1460-2075.
\newblock \doi{10.15252/EMBJ.201593561}.

\bibitem[Lepage and Junier(2019)]{Lepage2019Aug}
Thibaut Lepage and Ivan Junier.
\newblock {A polymer model of bacterial supercoiled DNA including structural transitions of the double helix}.
\newblock \emph{Physica A}, 527:\penalty0 121196, August 2019.
\newblock ISSN 0378-4371.
\newblock \doi{10.1016/j.physa.2019.121196}.

\bibitem[Lim et~al.(2014)Lim, Surovtsev, Beltran, Huang, Bewersdorf, and Jacobs-Wagner]{Lim2014May}
Hoong~Chuin Lim, Ivan~Vladimirovich Surovtsev, Bruno~Gabriel Beltran, Fang Huang, J{\ifmmode\ddot{o}\else\"{o}\fi}rg Bewersdorf, and Christine Jacobs-Wagner.
\newblock {Evidence for a DNA-relay mechanism in ParABS-mediated chromosome segregation}.
\newblock \emph{eLife}, May 2014.
\newblock \doi{10.7554/eLife.02758}.

\bibitem[Lioy et~al.(2018)Lioy, Cournac, Marbouty, Duigou, Mozziconacci, Esp{\ifmmode\acute{e}\else\'{e}\fi}li, Boccard, and Koszul]{Lioy2018Feb}
Virginia~S. Lioy, Axel Cournac, Martial Marbouty, St{\ifmmode\acute{e}\else\'{e}\fi}phane Duigou, Julien Mozziconacci, Olivier Esp{\ifmmode\acute{e}\else\'{e}\fi}li, Fr{\ifmmode\acute{e}\else\'{e}\fi}d{\ifmmode\acute{e}\else\'{e}\fi}ric Boccard, and Romain Koszul.
\newblock {Multiscale Structuring of the E. coli Chromosome by Nucleoid-Associated and Condensin Proteins}.
\newblock \emph{Cell}, 172\penalty0 (4):\penalty0 771--783.e18, February 2018.
\newblock ISSN 0092-8674.
\newblock \doi{10.1016/j.cell.2017.12.027}.

\bibitem[Lioy et~al.(2021)Lioy, Junier, and Boccard]{Lioy2021Oct}
Virginia~S. Lioy, Ivan Junier, and Fr{\ifmmode\acute{e}\else\'{e}\fi}d{\ifmmode\acute{e}\else\'{e}\fi}ric Boccard.
\newblock {Multiscale Dynamic Structuring of Bacterial Chromosomes}.
\newblock \emph{Annu. Rev. Microbiol.}, 75\penalty0 (1):\penalty0 541--561, October 2021.
\newblock ISSN 0066-4227.
\newblock \doi{10.1146/annurev-micro-033021-113232}.

\bibitem[Liu et~al.(2023)Liu, Qiu, Hua, and Ma]{Liu2023Oct}
Tong Liu, Qin-Tian Qiu, Kang-Jian Hua, and Bin-Guang Ma.
\newblock {Evaluation of chromosome structure modelling tools in bacteria}.
\newblock \emph{bioRxiv}, page 2023.10.26.564237, October 2023.

\bibitem[Marbouty et~al.(2015)Marbouty, Gall, Cattoni, Cournac, Koh, Fiche, Mozziconacci, Murray, Koszul, and Nollmann]{Marbouty2015}
Martial Marbouty, Antoine~Le Gall, Diego~I. Cattoni, Axel Cournac, Alan Koh, Jean~Bernard Fiche, Julien Mozziconacci, Heath Murray, Romain Koszul, and Marcelo Nollmann.
\newblock { Condensin- and Replication-Mediated Bacterial Chromosome Folding and Origin Condensation Revealed by Hi-C and Super-resolution Imaging }.
\newblock \emph{Molecular Cell}, 59:\penalty0 588--602, 8 2015.
\newblock ISSN 1097-2765.
\newblock \doi{10.1016/J.MOLCEL.2015.07.020}.

\bibitem[Marti-Renom et~al.(2018)Marti-Renom, Almouzni, Bickmore, Bystricky, Cavalli, Fraser, Gasser, Giorgetti, Heard, Nicodemi, Nollmann, Orozco, Pombo, and Torres-Padilla]{Marti-Renom2018Oct}
Marc~A. Marti-Renom, Genevieve Almouzni, Wendy~A. Bickmore, Kerstin Bystricky, Giacomo Cavalli, Peter Fraser, Susan~M. Gasser, Luca Giorgetti, Edith Heard, Mario Nicodemi, Marcelo Nollmann, Modesto Orozco, Ana Pombo, and Maria-Elena Torres-Padilla.
\newblock {Challenges and guidelines toward 4D nucleome data and model standards}.
\newblock \emph{Nat. Genet.}, 50:\penalty0 1352--1358, October 2018.
\newblock ISSN 1546-1718.
\newblock \doi{10.1038/s41588-018-0236-3}.

\bibitem[McCord et~al.(2020)McCord, Kaplan, and Giorgetti]{McCord2020}
Rachel~Patton McCord, Noam Kaplan, and Luca Giorgetti.
\newblock {Chromosome Conformation Capture and Beyond: Toward an Integrative View of Chromosome Structure and Function }.
\newblock \emph{Molecular cell}, 77:\penalty0 688--708, 2 2020.
\newblock ISSN 1097-4164.
\newblock \doi{10.1016/J.MOLCEL.2019.12.021}.

\bibitem[Messelink et~al.(2021)Messelink, van Teeseling, Janssen, Thanbichler, and Broedersz]{Messelink2021}
Joris~J.B. Messelink, Muriel~C.F. van Teeseling, Jacqueline Janssen, Martin Thanbichler, and Chase~P. Broedersz.
\newblock Learning the distribution of single-cell chromosome conformations in bacteria reveals emergent order across genomic scales.
\newblock \emph{Nature Communications}, 12:\penalty0 1--9, 3 2021.
\newblock ISSN 2041-1723.
\newblock \doi{10.1038/s41467-021-22189-x}.

\bibitem[Miangolarra et~al.(2021)Miangolarra, Li, Joanny, Wingreen, and Castellana]{Miangolarra2021Oct}
A.~Movilla Miangolarra, Sophia Hsin-Jung Li, Jean-Fran{\ifmmode\mbox{\c{c}}\else\c{c}\fi}ois Joanny, Ned~S. Wingreen, and Michele Castellana.
\newblock {Steric interactions and out-of-equilibrium processes control the internal organization of bacteria}.
\newblock \emph{Proc. Natl. Acad. Sci. U.S.A.}, 118\penalty0 (43):\penalty0 e2106014118, October 2021.
\newblock \doi{10.1073/pnas.2106014118}.

\bibitem[Miermans and Broedersz(2018)]{Miermans2018}
Christiaan~A. Miermans and Chase~P. Broedersz.
\newblock {Bacterial chromosome organization by collective dynamics of SMC condensins }.
\newblock \emph{Journal of The Royal Society Interface}, 15, 10 2018.
\newblock ISSN 17425662.
\newblock \doi{10.1098/RSIF.2018.0495}.

\bibitem[Minina and Arnold(2014)]{Minina2014}
Elena Minina and Axel Arnold.
\newblock {Induction of entropic segregation: the first step is the hardest}.
\newblock \emph{Soft Matter}, 10\penalty0 (31):\penalty0 5836--5841, 2014.
\newblock \doi{10.1039/C4SM00286E}.

\bibitem[Minina and Arnold(2015)]{Minina2015Jul}
Elena Minina and Axel Arnold.
\newblock {Entropic Segregation of Ring Polymers in Cylindrical Confinement}.
\newblock \emph{Macromolecules}, 48\penalty0 (14):\penalty0 4998--5005, July 2015.
\newblock ISSN 0024-9297.
\newblock \doi{10.1021/acs.macromol.5b00636}.

\bibitem[Mitra et~al.(2022{\natexlab{a}})Mitra, Pande, and Chatterji]{Mitra2022July}
Debarshi Mitra, Shreerang Pande, and Apratim Chatterji.
\newblock {Polymer architecture orchestrates the segregation and spatial organization of replicating E. coli chromosomes in slow growth}.
\newblock \emph{Soft Matter}, 18\penalty0 (30):\penalty0 5615--5631, 2022{\natexlab{a}}.
\newblock \doi{10.1039/D2SM00734G}.

\bibitem[Mitra et~al.(2022{\natexlab{b}})Mitra, Pande, and Chatterji]{Mitra2022Nov}
Debarshi Mitra, Shreerang Pande, and Apratim Chatterji.
\newblock {Topology-driven spatial organization of ring polymers under confinement}.
\newblock \emph{Phys. Rev. E}, 106\penalty0 (5):\penalty0 054502, November 2022{\natexlab{b}}.
\newblock ISSN 2470-0053.
\newblock \doi{10.1103/PhysRevE.106.054502}.

\bibitem[Murray et~al.(2006)Murray, Ferreira, and Errington]{Murray2006Sep}
Heath Murray, Henrique Ferreira, and Jeff Errington.
\newblock {The bacterial chromosome segregation protein Spo0J spreads along DNA from parS nucleation sites}.
\newblock \emph{Mol. Microbiol.}, 61\penalty0 (5):\penalty0 1352--1361, September 2006.
\newblock ISSN 0950-382X.
\newblock \doi{10.1111/j.1365-2958.2006.05316.x}.

\bibitem[Murray and Sourjik(2017)]{Murray2017Oct}
Se{\ifmmode\acute{a}\else\'{a}\fi}n~M. Murray and Victor Sourjik.
\newblock {Self-organization and positioning of bacterial protein clusters}.
\newblock \emph{Nat. Phys.}, 13:\penalty0 1006--1013, October 2017.
\newblock ISSN 1745-2481.
\newblock \doi{10.1038/nphys4155}.

\bibitem[Mäkelä and Sherratt(2020)]{Makela2020}
Jarno Mäkelä and David~J. Sherratt.
\newblock {Organization of the Escherichia coli Chromosome by a MukBEF Axial Core}.
\newblock \emph{Molecular Cell}, 78:\penalty0 250--260.e5, 4 2020.
\newblock ISSN 1097-2765.
\newblock \doi{10.1016/J.MOLCEL.2020.02.003}.

\bibitem[Oluwadare et~al.(2019)Oluwadare, Highsmith, and Cheng]{Oluwadare2019Dec}
Oluwatosin Oluwadare, Max Highsmith, and Jianlin Cheng.
\newblock {An Overview of Methods for Reconstructing 3-D Chromosome and Genome Structures from Hi-C Data}.
\newblock \emph{Biol. Proced. Online}, 21\penalty0 (1):\penalty0 1--20, December 2019.
\newblock ISSN 1480-9222.
\newblock \doi{10.1186/s12575-019-0094-0}.

\bibitem[Osorio-Valeriano et~al.(2021)Osorio-Valeriano, Altegoer, Das, Steinchen, Panis, Connolley, Giacomelli, Feddersen, Corrales-Guerrero, Giammarinaro, Han{\ss}mann, Bramkamp, Viollier, Murray, Sch{\ifmmode\ddot{a}\else\"{a}\fi}fer, Bange, and Thanbichler]{Osorio-Valeriano2021Oct}
Manuel Osorio-Valeriano, Florian Altegoer, Chandan~K. Das, Wieland Steinchen, Ga{\ifmmode\ddot{e}\else\"{e}\fi}l Panis, Lara Connolley, Giacomo Giacomelli, Helge Feddersen, Laura Corrales-Guerrero, Pietro~I. Giammarinaro, Juri Han{\ss}mann, Marc Bramkamp, Patrick~H. Viollier, Se{\ifmmode\acute{a}\else\'{a}\fi}n Murray, Lars~V. Sch{\ifmmode\ddot{a}\else\"{a}\fi}fer, Gert Bange, and Martin Thanbichler.
\newblock {The CTPase activity of ParB determines the size and dynamics of prokaryotic DNA partition complexes}.
\newblock \emph{Mol. Cell}, 81\penalty0 (19):\penalty0 3992--4007.e10, October 2021.
\newblock ISSN 1097-2765.
\newblock \doi{10.1016/j.molcel.2021.09.004}.

\bibitem[Pinho et~al.(2013)Pinho, Kjos, and Veening]{Pinho2013}
Mariana~G. Pinho, Morten Kjos, and Jan~Willem Veening.
\newblock How to get (a)round: mechanisms controlling growth and division of coccoid bacteria.
\newblock \emph{Nature Reviews Microbiology 2013 11:9}, 11:\penalty0 601--614, 8 2013.
\newblock ISSN 1740-1534.
\newblock \doi{10.1038/nrmicro3088}.

\bibitem[Polson and Kerry(2018)]{Polson2018Aug}
James~M. Polson and Deanna R.-M. Kerry.
\newblock {Segregation of polymers under cylindrical confinement: effects of polymer topology and crowding}.
\newblock \emph{Soft Matter}, 14\penalty0 (30):\penalty0 6360--6373, August 2018.
\newblock ISSN 1744-683X.
\newblock \doi{10.1039/C8SM01062E}.

\bibitem[Polson and Zhu(2021)]{Polson2021Jan}
James~M. Polson and Qinxin Zhu.
\newblock {Free energy and segregation dynamics of two channel-confined polymers of different lengths}.
\newblock \emph{Phys. Rev. E}, 103\penalty0 (1):\penalty0 012501, January 2021.
\newblock \doi{10.1103/PhysRevE.103.012501}.

\bibitem[Ren et~al.(2022)Ren, Liao, Karaboja, Barton, Schantz, Mejia-Santana, Fuqua, and Wang]{Ren2022Feb}
Zhongqing Ren, Qin Liao, Xheni Karaboja, Ian~S. Barton, Eli~G. Schantz, Adrian Mejia-Santana, Clay Fuqua, and Xindan Wang.
\newblock {Conformation and dynamic interactions of the multipartite genome in Agrobacterium tumefaciens}.
\newblock \emph{Proc. Natl. Acad. Sci. U.S.A.}, 119\penalty0 (6):\penalty0 e2115854119, February 2022.
\newblock \doi{10.1073/pnas.2115854119}.

\bibitem[Rivas and Minton(2016)]{Rivas2016}
Germán Rivas and Allen~P. Minton.
\newblock { Macromolecular Crowding In Vitro, In Vivo, and In Between }.
\newblock \emph{Trends in Biochemical Sciences}, 41:\penalty0 970--981, 11 2016.
\newblock ISSN 13624326.
\newblock \doi{10.1016/j.tibs.2016.08.013}.

\bibitem[Roggiani and Goulian(2015)]{Roggiani2015Nov}
Manuela Roggiani and Mark Goulian.
\newblock {Chromosome-Membrane Interactions in Bacteria}.
\newblock \emph{Annu. Rev. Genet.}, 49\penalty0 (1):\penalty0 115--129, November 2015.
\newblock ISSN 0066-4197.
\newblock \doi{10.1146/annurev-genet-112414-054958}.

\bibitem[Russel et~al.(2012)Russel, Lasker, Webb, Vel{\ifmmode\acute{a}\else\'{a}\fi}zquez-Muriel, Tjioe, Schneidman-Duhovny, Peterson, and Sali]{Russel2012Jan}
Daniel Russel, Keren Lasker, Ben Webb, Javier Vel{\ifmmode\acute{a}\else\'{a}\fi}zquez-Muriel, Elina Tjioe, Dina Schneidman-Duhovny, Bret Peterson, and Andrej Sali.
\newblock {Putting the Pieces Together: Integrative Modeling Platform Software for Structure Determination of Macromolecular Assemblies}.
\newblock \emph{PLoS Biol.}, 10\penalty0 (1):\penalty0 e1001244, January 2012.
\newblock ISSN 1545-7885.
\newblock \doi{10.1371/journal.pbio.1001244}.

\bibitem[Sanchez et~al.(2015)Sanchez, Cattoni, Walter, Rech, Parmeggiani, Nollmann, and Bouet]{Sanchez2015Aug}
Aurore Sanchez, Diego~I. Cattoni, Jean-Charles Walter, J{\ifmmode\acute{e}\else\'{e}\fi}r{\ifmmode\hat{o}\else\^{o}\fi}me Rech, Andrea Parmeggiani, Marcelo Nollmann, and Jean-Yves Bouet.
\newblock {Stochastic Self-Assembly of ParB Proteins Builds the Bacterial DNA Segregation Apparatus}.
\newblock \emph{Cell Systems}, 1\penalty0 (2):\penalty0 163--173, August 2015.
\newblock ISSN 2405-4712.
\newblock \doi{10.1016/j.cels.2015.07.013}.

\bibitem[Song and Loparo(2015)]{Song2015}
Dan Song and Joseph~J. Loparo.
\newblock Building bridges within the bacterial chromosome.
\newblock \emph{Trends in Genetics}, 31:\penalty0 164--173, 3 2015.
\newblock ISSN 0168-9525.
\newblock \doi{10.1016/J.TIG.2015.01.003}.

\bibitem[Spahn et~al.(2023)Spahn, Middlemiss, G{\ifmmode\acute{o}\else\'{o}\fi}mez-de Mariscal, Henriques, Bode, Holden, and Heilemann]{Spahn2023Oct}
Christoph Spahn, Stuart Middlemiss, Estibaliz G{\ifmmode\acute{o}\else\'{o}\fi}mez-de Mariscal, Ricardo Henriques, Helge~B. Bode, S{\ifmmode\acute{e}\else\'{e}\fi}amus Holden, and Mike Heilemann.
\newblock {Transertion and cell geometry organize the Escherichia coli nucleoid during rapid growth}.
\newblock \emph{bioRxiv}, page 2023.10.16.562172, October 2023.

\bibitem[Stevens et~al.(2023)Stevens, Gr{\ifmmode\ddot{u}\else\"{u}\fi}newald, van Tilburg, K{\ifmmode\ddot{o}\else\"{o}\fi}nig, Gilbert, Brier, Thornburg, Luthey-Schulten, and Marrink]{Stevens2023}
Jan~A. Stevens, Fabian Gr{\ifmmode\ddot{u}\else\"{u}\fi}newald, P.~A.~Marco van Tilburg, Melanie K{\ifmmode\ddot{o}\else\"{o}\fi}nig, Benjamin~R. Gilbert, Troy~A. Brier, Zane~R. Thornburg, Zaida Luthey-Schulten, and Siewert~J. Marrink.
\newblock {Molecular dynamics simulation of an entire cell}.
\newblock \emph{Front. Chem.}, 11:\penalty0 1106495, January 2023.
\newblock ISSN 2296-2646.
\newblock \doi{10.3389/fchem.2023.1106495}.

\bibitem[Subramanian and Murray(2023)]{Subramanian2023Apr}
Srikanth Subramanian and Se{\ifmmode\acute{a}\else\'{a}\fi}n~M. Murray.
\newblock {Subdiffusive movement of chromosomal loci in bacteria explained by DNA bridging}.
\newblock \emph{Phys. Rev. Res.}, 5\penalty0 (2):\penalty0 023034, April 2023.
\newblock ISSN 2643-1564.
\newblock \doi{10.1103/PhysRevResearch.5.023034}.

\bibitem[Surovtsev et~al.(2016)Surovtsev, Campos, and Jacobs-Wagner]{Surovtsev2016Nov}
Ivan~V. Surovtsev, Manuel Campos, and Christine Jacobs-Wagner.
\newblock {DNA-relay mechanism is sufficient to explain ParA-dependent intracellular transport and patterning of single and multiple cargos}.
\newblock \emph{Proc. Natl. Acad. Sci. U.S.A.}, 113\penalty0 (46):\penalty0 E7268--E7276, November 2016.
\newblock \doi{10.1073/pnas.1616118113}.

\bibitem[Swain et~al.(2019)Swain, Mulder, and Chaudhuri]{Swain2019}
Pinaki Swain, Bela~M. Mulder, and Debasish Chaudhuri.
\newblock Confinement and crowding control the morphology and dynamics of a model bacterial chromosome.
\newblock \emph{Soft Matter}, 15:\penalty0 2677--2687, 3 2019.
\newblock ISSN 17446848.
\newblock \doi{10.1039/C8SM02092B}.

\bibitem[Ti{\ifmmode\check{s}\else\v{s}\fi}ma et~al.(2022)Ti{\ifmmode\check{s}\else\v{s}\fi}ma, Panoukidou, Antar, Soh, Barth, Pradhan, Barth, van~der Torre, Michieletto, Gruber, and Dekker]{Tisma2022Jun}
Milo{\ifmmode\check{s}\else\v{s}\fi} Ti{\ifmmode\check{s}\else\v{s}\fi}ma, Maria Panoukidou, Hammam Antar, Young-Min Soh, Roman Barth, Biswajit Pradhan, Anders Barth, Jaco van~der Torre, Davide Michieletto, Stephan Gruber, and Cees Dekker.
\newblock {ParB proteins can bypass DNA-bound roadblocks via dimer-dimer recruitment}.
\newblock \emph{Sci. Adv.}, 8\penalty0 (26), June 2022.
\newblock ISSN 2375-2548.
\newblock \doi{10.1126/sciadv.abn3299}.

\bibitem[Ti{\ifmmode\check{s}\else\v{s}\fi}ma et~al.(2023)Ti{\ifmmode\check{s}\else\v{s}\fi}ma, Janissen, Antar, Martin-Gonzalez, Barth, Beekman, van~der Torre, Michieletto, Gruber, and Dekker]{Tisma2023Nov}
Milo{\ifmmode\check{s}\else\v{s}\fi} Ti{\ifmmode\check{s}\else\v{s}\fi}ma, Richard Janissen, Hammam Antar, Alejandro Martin-Gonzalez, Roman Barth, Twan Beekman, Jaco van~der Torre, Davide Michieletto, Stephan Gruber, and Cees Dekker.
\newblock {Dynamic ParB{\textendash}DNA interactions initiate and maintain a partition condensate for bacterial chromosome segregation}.
\newblock \emph{Nucleic Acids Res.}, 51\penalty0 (21):\penalty0 11856--11875, November 2023.
\newblock ISSN 0305-1048.
\newblock \doi{10.1093/nar/gkad868}.

\bibitem[Umbarger et~al.(2011)Umbarger, Toro, Wright, Porreca, Ba{\ifmmode\grave{u}\else\`{u}\fi}, Hong, Fero, Zhu, Marti-Renom, McAdams, Shapiro, Dekker, and Church]{Umbarger2011Oct}
Mark~A. Umbarger, Esteban Toro, Matthew~A. Wright, Gregory~J. Porreca, Davide Ba{\ifmmode\grave{u}\else\`{u}\fi}, Sun-Hae Hong, Michael~J. Fero, Lihua~J. Zhu, Marc~A. Marti-Renom, Harley~H. McAdams, Lucy Shapiro, Job Dekker, and George~M. Church.
\newblock {The Three-Dimensional Architecture of a Bacterial Genome and Its Alteration by Genetic Perturbation}.
\newblock \emph{Mol. Cell}, 44\penalty0 (2):\penalty0 252--264, October 2011.
\newblock ISSN 1097-2765.
\newblock \doi{10.1016/j.molcel.2011.09.010}.

\bibitem[Viollier et~al.(2004)Viollier, Thanbichler, McGrath, West, Meewan, McAdams, and Shapiro]{Viollier2004Jun}
Patrick~H. Viollier, Martin Thanbichler, Patrick~T. McGrath, Lisandra West, Maliwan Meewan, Harley~H. McAdams, and Lucy Shapiro.
\newblock {Rapid and sequential movement of individual chromosomal loci to specific subcellular locations during bacterial DNA replication}.
\newblock \emph{Proc. Natl. Acad. Sci. U.S.A.}, 101\penalty0 (25):\penalty0 9257--9262, June 2004.
\newblock \doi{10.1073/pnas.0402606101}.

\bibitem[Walter et~al.(2017)Walter, Dorignac, Lorman, Rech, Bouet, Nollmann, Palmeri, Parmeggiani, and Geniet]{Walter2017Jul}
J.-C. Walter, J.~Dorignac, V.~Lorman, J.~Rech, J.-Y. Bouet, M.~Nollmann, J.~Palmeri, A.~Parmeggiani, and F.~Geniet.
\newblock {Surfing on Protein Waves: Proteophoresis as a Mechanism for Bacterial Genome Partitioning}.
\newblock \emph{Phys. Rev. Lett.}, 119\penalty0 (2):\penalty0 028101, July 2017.
\newblock ISSN 1079-7114.
\newblock \doi{10.1103/PhysRevLett.119.028101}.

\bibitem[Walter et~al.(2021)Walter, Lepage, Dorignac, Geniet, Parmeggiani, Palmeri, Bouet, and Junier]{Walter2021Apr}
Jean-Charles Walter, Thibaut Lepage, J{\ifmmode\acute{e}\else\'{e}\fi}r{\ifmmode\hat{o}\else\^{o}\fi}me Dorignac, Fr{\ifmmode\acute{e}\else\'{e}\fi}d{\ifmmode\acute{e}\else\'{e}\fi}ric Geniet, Andrea Parmeggiani, John Palmeri, Jean-Yves Bouet, and Ivan Junier.
\newblock {Supercoiled DNA and non-equilibrium formation of protein complexes: A quantitative model of the nucleoprotein ParBS partition complex}.
\newblock \emph{PLoS Comput. Biol.}, 17\penalty0 (4):\penalty0 e1008869, April 2021.
\newblock ISSN 1553-7358.
\newblock \doi{10.1371/journal.pcbi.1008869}.

\bibitem[Wang et~al.(2014)Wang, Montero~Llopis, and Rudner]{Wang2014Sep}
Xindan Wang, Paula Montero~Llopis, and David~Z. Rudner.
\newblock {Bacillus subtilis chromosome organization oscillates between two distinct patterns}.
\newblock \emph{Proc. Natl. Acad. Sci. U.S.A.}, 111\penalty0 (35):\penalty0 12877--12882, September 2014.
\newblock \doi{10.1073/pnas.1407461111}.

\bibitem[Wang et~al.(2015)Wang, Le, Lajoie, Dekker, Laub, and Rudner]{Wang2015}
Xindan Wang, Tung~B.K. Le, Bryan~R. Lajoie, Job Dekker, Michael~T. Laub, and David~Z. Rudner.
\newblock { Condensin promotes the juxtaposition of DNA flanking its loading site in Bacillus subtilis }.
\newblock \emph{Genes \& Development}, 29:\penalty0 1661--1675, 8 2015.
\newblock ISSN 0890-9369.
\newblock \doi{10.1101/GAD.265876.115}.

\bibitem[Wang et~al.(2017)Wang, Brand{\ifmmode\tilde{a}\else\~{a}\fi}o, Le, Laub, and Rudner]{Wang2017}
Xindan Wang, Hugo~B. Brand{\ifmmode\tilde{a}\else\~{a}\fi}o, Tung B.~K. Le, Michael~T. Laub, and David~Z. Rudner.
\newblock {Bacillus subtilis SMC complexes juxtapose chromosome arms as they travel from origin to terminus}.
\newblock \emph{Science}, 355\penalty0 (6324):\penalty0 524--527, February 2017.
\newblock ISSN 1095-9203.
\newblock \doi{10.1126/science.aai8982}.

\bibitem[Wasim et~al.(2021)Wasim, Gupta, and Mondal]{Wasim2021}
Abdul Wasim, Ankit Gupta, and Jagannath Mondal.
\newblock { A Hi–C data-integrated model elucidates E. coli chromosome’s multiscale organization at various replication stages }.
\newblock \emph{Nucleic Acids Research}, 49:\penalty0 3077--3091, 4 2021.
\newblock ISSN 0305-1048.
\newblock \doi{10.1093/NAR/GKAB094}.

\bibitem[Wasim et~al.(2023{\natexlab{a}})Wasim, Bera, and Mondal]{Wasim2023Apr}
Abdul Wasim, Palash Bera, and Jagannath Mondal.
\newblock {Development of a Data-Driven Integrative Model of a Bacterial Chromosome}.
\newblock \emph{J. Chem. Theory Comput.}, 2023, April 2023{\natexlab{a}}.
\newblock ISSN 1549-9618.
\newblock \doi{10.1021/acs.jctc.3c00118}.

\bibitem[Wasim et~al.(2023{\natexlab{b}})Wasim, Bera, and Mondal]{Wasim2023Jul}
Abdul Wasim, Palash Bera, and Jagannath Mondal.
\newblock {On the Spatial Positioning of Ribosomes around chromosome in E. coli Cytoplasm}.
\newblock \emph{bioRxiv}, page 2023.07.04.547709, July 2023{\natexlab{b}}.

\bibitem[Wasim et~al.(2023{\natexlab{c}})Wasim, Gupta, Bera, and Mondal]{Wasim2023}
Abdul Wasim, Ankit Gupta, Palash Bera, and Jagannath Mondal.
\newblock { Interpretation of organizational role of proteins on E. coli nucleoid via Hi-C integrated model }.
\newblock \emph{Biophysical journal}, 122:\penalty0 63--81, 1 2023{\natexlab{c}}.
\newblock ISSN 1542-0086.
\newblock \doi{10.1016/J.BPJ.2022.11.2938}.

\bibitem[Wilhelm et~al.(2015)Wilhelm, B{\ifmmode\ddot{u}\else\"{u}\fi}rmann, Minnen, Shin, Toseland, Oh, and Gruber]{Wilhelm2015May}
Larissa Wilhelm, Frank B{\ifmmode\ddot{u}\else\"{u}\fi}rmann, Anita Minnen, Ho-Chul Shin, Christopher~P. Toseland, Byung-Ha Oh, and Stephan Gruber.
\newblock {SMC condensin entraps chromosomal DNA by an ATP hydrolysis dependent loading mechanism in Bacillus subtilis}.
\newblock \emph{eLife}, May 2015.
\newblock \doi{10.7554/eLife.06659}.

\bibitem[Xiang et~al.(2021)Xiang, Surovtsev, Chang, Govers, Parry, Liu, and Jacobs-Wagner]{Xiang2021Jul}
Yingjie Xiang, Ivan~V. Surovtsev, Yunjie Chang, Sander~K. Govers, Bradley~R. Parry, Jun Liu, and Christine Jacobs-Wagner.
\newblock {Interconnecting solvent quality, transcription, and chromosome folding in Escherichia coli}.
\newblock \emph{Cell}, 184\penalty0 (14):\penalty0 3626--3642.e14, July 2021.
\newblock ISSN 0092-8674.
\newblock \doi{10.1016/j.cell.2021.05.037}.

\bibitem[Y{\ifmmode\acute{a}\else\'{a}\fi}{\ifmmode\tilde{n}\else\~{n}\fi}ez-Cuna and Koszul(2023)]{Yanez-Cuna2023Oct}
Fares~Osam Y{\ifmmode\acute{a}\else\'{a}\fi}{\ifmmode\tilde{n}\else\~{n}\fi}ez-Cuna and Romain Koszul.
\newblock {Insights in bacterial genome folding}.
\newblock \emph{Curr. Opin. Struct. Biol.}, 82:102679., October 2023.
\newblock ISSN 1879-033X.
\newblock \doi{10.1016/j.sbi.2023.102679}.

\bibitem[Yildirim and Feig(2018)]{Yildirim2018May}
Asli Yildirim and Michael Feig.
\newblock {High-resolution 3D models of Caulobacter crescentus chromosome reveal genome structural variability and organization}.
\newblock \emph{Nucleic Acids Res.}, 46\penalty0 (8):\penalty0 3937, May 2018.
\newblock \doi{10.1093/nar/gky141}.

\bibitem[Yu et~al.(2021)Yu, Wu, Meng, Chu, Li, and Wu]{Yu2021}
Shi Yu, Jiaxin Wu, Xianliang Meng, Ruizhi Chu, Xiao Li, and Guoguang Wu.
\newblock { Mesoscale Simulation of Bacterial Chromosome and Cytoplasmic Nanoparticles in Confinement }.
\newblock \emph{Entropy 2021, Vol. 23, Page 542}, 23:\penalty0 542, 4 2021.
\newblock ISSN 1099-4300.
\newblock \doi{10.3390/E23050542}.

\end{thebibliography}

\end{document}